\documentclass[%
 reprint,
superscriptaddress,
nofootinbib,
 amsmath,amssymb,
 aps,
prd,
]{revtex4-2}

\usepackage{graphicx}
\usepackage{dcolumn}
\usepackage{bm}
\usepackage{lipsum}
\usepackage{xfrac}

\usepackage{hyperref}
\usepackage{amssymb}
\usepackage{xcolor}

\begin{document}


\title{Quartz fluorescence backgrounds in xenon particle detectors}

\author{P. Sorensen}\email{pfsorensen@lbl.gov}
\affiliation{Lawrence Berkeley National Laboratory, 1 Cyclotron Rd, Berkeley, CA 94720, USA}
\author{R. Gibbons}
\affiliation{Lawrence Berkeley National Laboratory, 1 Cyclotron Rd, Berkeley, CA 94720, USA}
\affiliation{University of California, Berkeley, Department of Physics, Berkeley, CA 94720, USA}

\date{\today}

\begin{abstract}
It has been known for almost a decade that delayed photon noise with a power law time profile follows scintillation pulses in liquid xenon particle detectors. The origin of the noise has remained unknown, and in the past two years, has become an overwhelming background for low-threshold dark matter searches aimed at $\mathcal{O}(10)$~GeV dark matter particle masses, as well as measurements of coherent neutrino-nucleus scattering of \textsuperscript{8}B solar neutrinos. We have performed a comprehensive series of tests in a small liquid xenon cell at LBL, from which we conclude that the dominant component of this delayed photon noise is due to UV-induced fluorescence of quartz photosensor windows.
\end{abstract}

\maketitle

\section{Introduction}
Astrophysical observations and cosmological data indicate the existence of non-luminous, non-baryonic massive dark matter~\cite{Planck:2018vyg,Sofue:2000jx}. 
A favored candidate for this dark matter has been a hypothetical class of beyond-the-Standard-Model particles called Weakly Interacting Massive Particles (WIMPs) with typical masses in the range of 100 GeV/c\textsuperscript{2}~\cite{Schumann:2019eaa}. The most sensitive experiments searching for these particles use liquid xenon as a scattering target~\cite{LZ:2024zvo,XENON:2025vwd,PandaX-4T:2021bab}, deployed in a dual-phase Time Projection Chamber (TPC) \cite{Baudis:2023pzu}. 

Particle interaction events in this class of TPC consist of a primary scintillation pulse of 175~nm ultraviolet photons (S1), followed by electron drift, electron emission into the vapor, and an electron-induced proportional electroluminescence pulse of 175~nm ultraviolet photons (S2). The S2 signal occurs anywhere up to about a millisecond later, depending on the distance the electrons drift. The amplitude of a typical S2 pulse is a detector-dependent $\bar{a}\sim10^6$ photons for a 1 MeV electron interaction, considering 45 electrons/keV \cite{Szydagis:2022ikv} and an S2 gain factor of $\times25$ photons detected per electron. The dominant scintillation decay time is $\tau=27$~ns \cite{Abe_2018}, and S1 pulses of interest can be as small as two single photons in a few hundred ns coincidence window. 

In such detectors, it is known that ionizing events due to particle interactions are followed by a trickle of single electron signals extending in time for tens to hundreds of milliseconds after the originating event~\cite{XENON10:2011prx}. This delayed electron noise presents a limiting background for sub-GeV-mass dark matter searches in this class of detector \cite{Essig:2012yx}. Delayed electron noise was later found to be accompanied by delayed photon noise \cite{Sorensen:2017kpl}, both of which exhibit power law decay \cite{LUX:2020vbj}. The delayed photon noise has been hypothesized to originate from PTFE fluorescence \cite{Sorensen:2017kpl,LUX:2020vbj}, although dedicated measurements found no such effect \cite{Araujo_2019}. We write the delayed photon noise as $d_p(t) = \alpha\bar{a}t^k$ with $\alpha$ expected to be constant and $\bar{a}$ the average size of the excitation pulse of UV photons. An investigation of $d_p(t)$ in LZ found $\alpha=0.42$ and $k=-1.3$ \cite{Anderson-thesis}. 

Accidental coincidence (often referred to as AC) backgrounds $-$ partially caused by pile-up of the delayed photon noise $-$ have recently become dominant in searches for low-mass dark matter candidates \cite{Kaplan:2009ag,López-Fogliani_2021,Wang_2021} and observation of \textsuperscript{8}B solar neutrinos \cite{PandaX:2022aac,XENON:2024ijk}. This is a reflection of the ultra-low radiogenic backgrounds these experiments have achieved. It is also somewhat striking, because in the planning phases of these experiments, accidental coincidence backgrounds were either not mentioned \cite{XENON:2024wpa}, or were dismissed as sub-dominant \cite{Mount:2017qzi}. 

\begin{figure}[hb]
    \centering
    \includegraphics[width=0.49\textwidth]{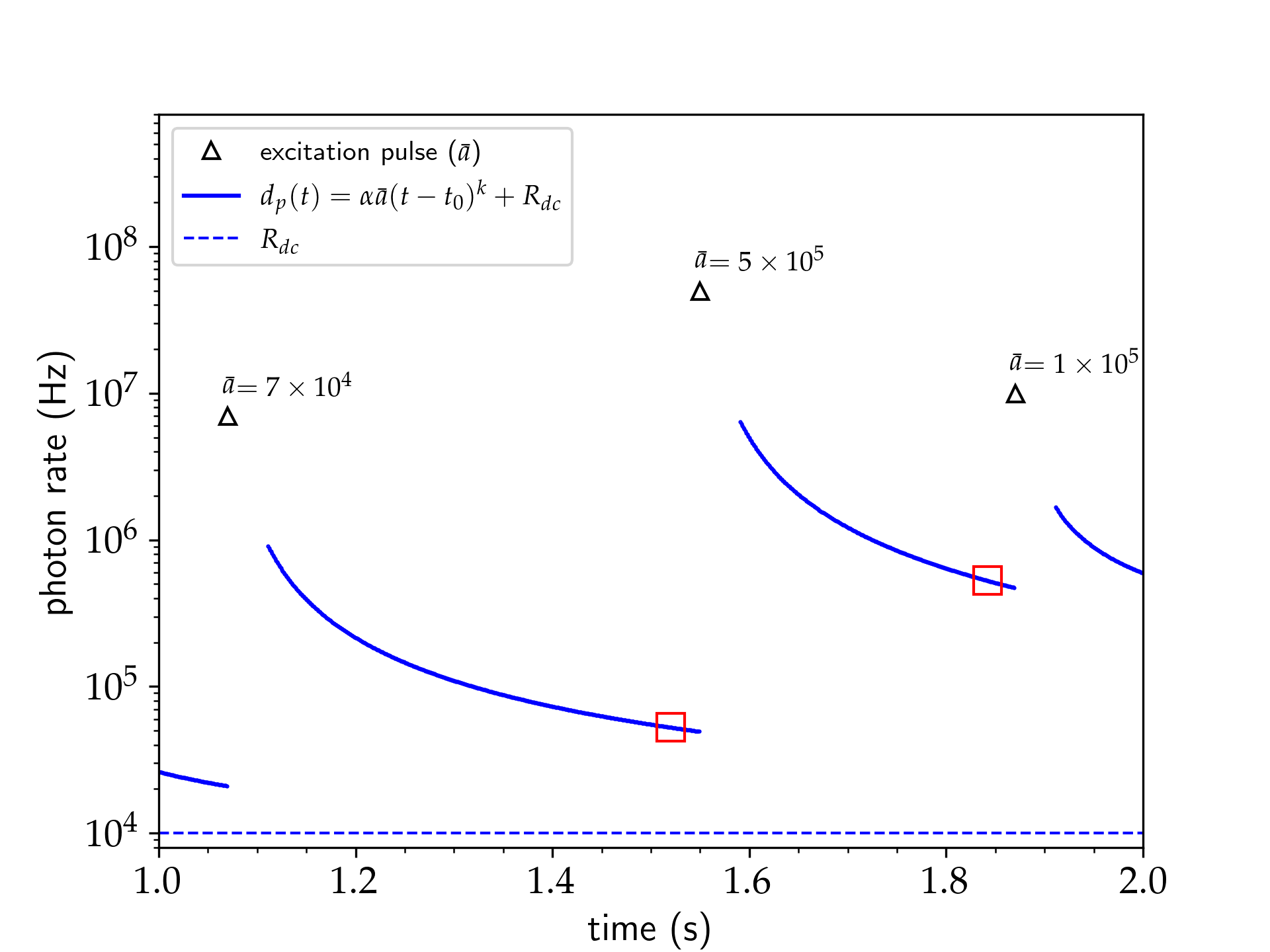}
    \caption{Simulated background events (triangles) in a hypothetical liquid xenon TPC dark matter search detector, and power law delayed photon rate $d_p(t)$. Red squares indicate regions with a factor $\times 1000$ difference in S1 accidental coincidence (AC) rate. }
   \label{fig:photon-rate}
\end{figure}

The problem is exemplified in Fig. \ref{fig:photon-rate}, showing simulated random background events (triangles) occurring at times $t_0$ in a hypothetical liquid xenon TPC. The total size of the S1 and S2 pulse in each background event constitutes the excitation pulse $\bar{a}$. The delayed photon noise is modeled as {\color{black} $d_p(t) = \alpha\bar{a}(t-t_0)^k +R_{dc}$, and shown with a hold-off time of 40~ms} in which $d_p(t)$ is not calculated. A non-zero hold-off is needed mathematically due to the $1/t$ nature of the noise, and is also typically applied experimentally to help mitigate the effects of the delayed photon noise. The total dark count rate $R_{dc}=10^4$~Hz could correspond to e.g. a detector with 250 PMTs, each with a dark count rate of 40~Hz. For this example, we chose {\color{black} $\alpha=0.2\times10^{3}$ and $k=-1.3$}, considering our measurements summarized in Table \ref{tab:alpha}. 

In an ideal dark matter search experiment, the photon counting rate would return to $R_{dc}$ shortly after each background event. However, the power law delayed photon noise perturbs the quiescent time between background events, during which experiments wish to look for small, candidate dark matter signals with S2 pulse sizes of {\color{black} $\sim1000$ photons}, and S1 pulse sizes of a handful of detected photons. In principle, a valid S1 pulse could be as small as two detected photons in two distinct photosensors ($n=2$), but such a signal has strong competition from accidental coincidence. The regions indicated by the red squares exemplify a factor $\times10$ difference in photon background rate. This increases the S1  $n=3$ AC probability in a 200~ns window from {\color{black} 5.8~Hz to 5.8~kHz}, as calculated from Eq. 4.1 of \cite{Gibbons:2023iux}. Rates for S1 $n=2$ AC backgrounds are a factor $\times250$ higher. Most of the search power reported in \cite{PandaX:2022aac,XENON:2024ijk} comes from such small S1 signals. Meanwhile, the power law nature of the noise coupled with an event rate of e.g., 5~Hz means that implementing a post-event time-trigger hold-off to reduce the photon rate is at odds with the necessity of maximizing search live time \cite{LZ:2024zvo}.  

A fundamental question has remained: \emph{what is the origin of the delayed photon noise?} We have investigated this issue with a 3~cm diameter cylindrical testbed at LBL and conclude that the majority of it is due to fluorescence of quartz\footnote{Optical grade SiO$_2$ is often referred to interchangeably in technical literature as fused silica, synthetic silica, or quartz.} photosensor windows, following exposure to UV photons from xenon scintillation.

\section{Experiment} 
We initially studied the delayed photon noise with the testbed operated as a TPC i.e., with S1 and S2 pulses \cite{Sorensen:2024idm}, in the style of the dark matter search experiments \cite{PandaX:2022aac,XENON:2024ijk}. These early studies utilized only the Hamamatsu S13371 sensors. In the TPC configuration, the delayed photon noise was measured follow a single power law time profile out to $t=5$~ms, and found to be completely insensitive to the xenon purity and the applied electric fields. We subsequently realized that power law delayed photon noise was also observed following the primary scintillation pulse (S1), with $\alpha$ consistent to within 20\%. The uncertainty in comparing $\alpha$ between the TPC and S1-only configurations is dominated by sensor saturation in the TPC case. The power law exponents were nearly identical in the TPC and S1-only cases, in the range $-1.3\lesssim k \lesssim -0.8$ depending on event selection. The present work is therefore restricted to the simpler case using only S1 as an excitation pulse, with no applied electric fields and no S2 pulses. 

To accurately count individual $\mathcal{O}(10)$~ns wide delayed single photons over a period of milliseconds following each excitation pulse, we employed a ``cascade trigger'': the initial 100~$\mu$s trigger window was automatically followed by three additional 100~$\mu$s windows, delayed by several hundred microseconds each. This allowed us to methodically sample the photons at later times, without loss of fidelity and within the capabilities of the data acquisition system. The trigger time of the excitation pulse $\bar{a}$ was set at 10~$\mu$s in the excitation pulse window, which allowed an additional data point to be acquired from the last 50~$\mu$s of the trigger window. 

\begin{figure}
    \centering
    \includegraphics[width=0.49\textwidth]{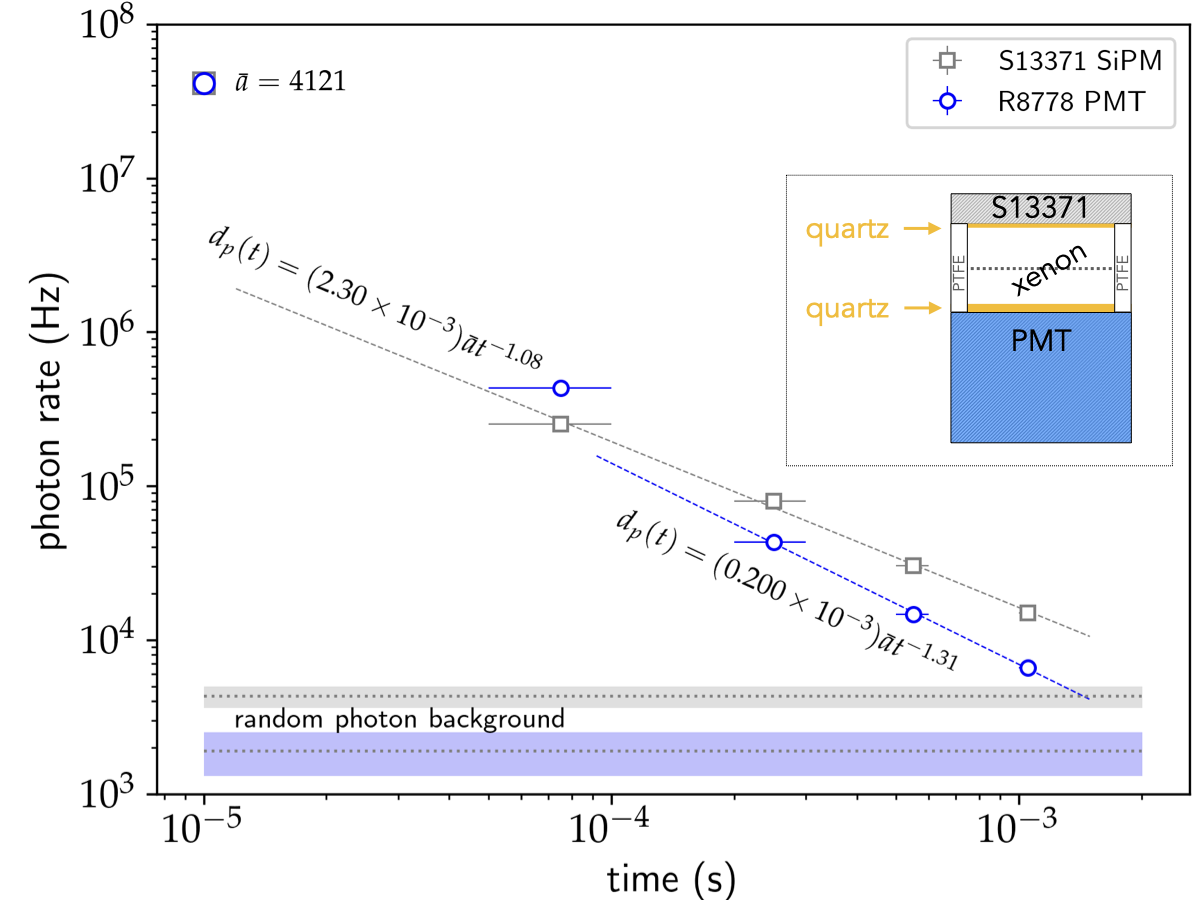}
    \caption{Measurement of the delayed photon rate following an excitation pulse $\bar{a}$ of UV photons from xenon scintillation. The experimental configuration is indicated inset.}
   \label{fig:xe}
\end{figure}

\subsection{Which material is fluorescing?}
A 2D schematic cross section of the cylindrical scintillation-only measurement apparatus is shown in Fig. \ref{fig:xe} (inset), indicating the single 5~cm diameter Hamamatsu R8778 PMT $-$ a sensor previously deployed in the LUX experiment \cite{LUX:2020vbj} $-$ and the array of four, four-channel Hamamatsu S13371 silicon photomultipliers (SiPM) viewing the active region filled with liquid xenon. The four corner channels were not read out, so signals from this array consisted of twelve 6~mm x 6~mm sensors. Walls were constructed of PTFE, with several 1~mm diameter radial holes (not shown) allowing xenon to flow into and out of the active region. A single high-transparency stainless steel gridded electrode bifurcated the region, and a $^{210}$Po alpha particle source plated in the center of the electrode caused xenon scintillation pulses at a rate of about 10~Hz. The testbed was maintained at a constant temperature $T=(-99 \pm 1)$~$^{\circ}$C, with a calibration uncertainty of about 1~C. The average response of many such measurements of $d_p(t)$ are shown in Fig. \ref{fig:xe}, following an average excitation pulse of $\bar{a}=4121$~photons detected. 

\begin{figure}[ht]
    \centering
    \includegraphics[width=0.49\textwidth]{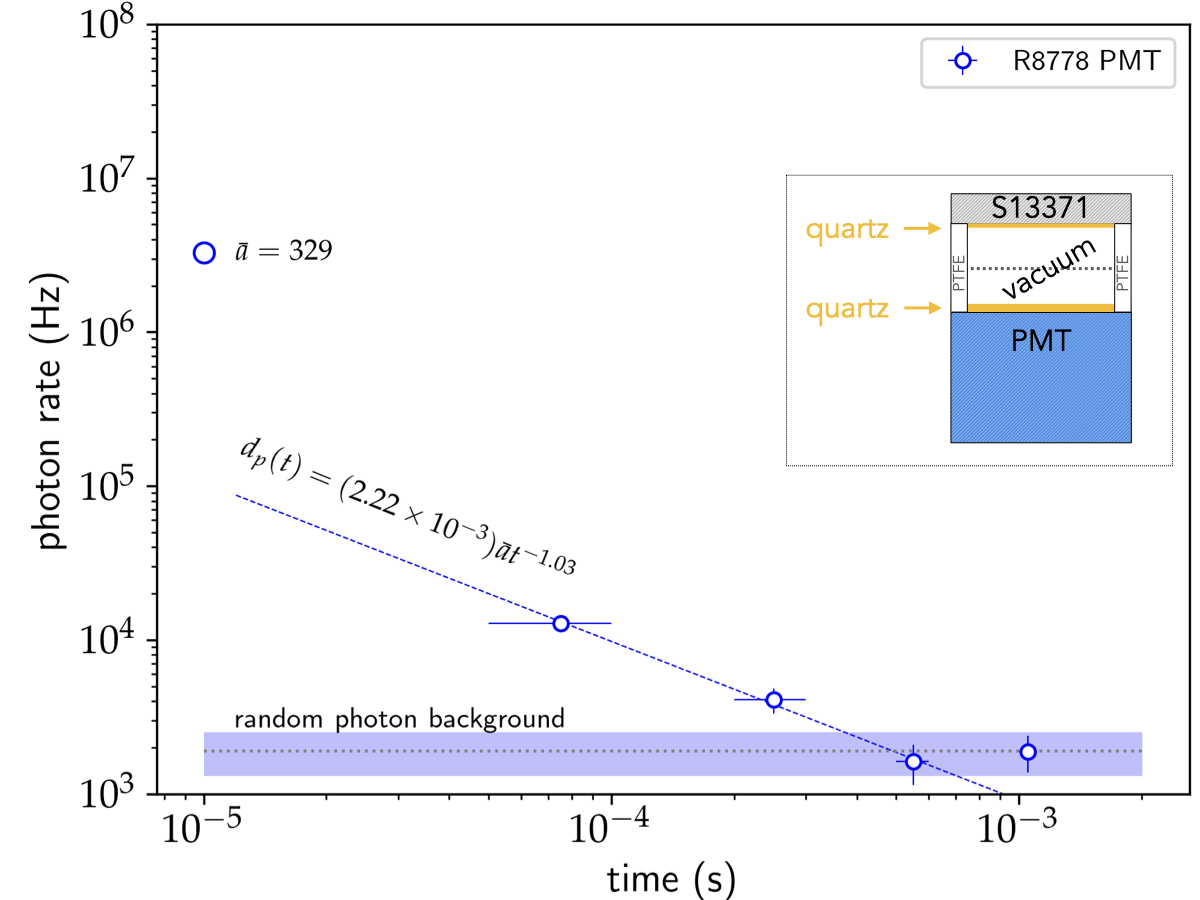}
    \caption{Measurement of the delayed photon rate following an excitation pulse $\bar{a}$ of UV photons from Cherenkov radiation. The experimental configuration is indicated inset.}
   \label{fig:vac}
\end{figure}

As a side note, we observe that for PMT signals greater than a few hundred photons, the PMT exhibits increased the photon counts in the $50-100~\mu$s region. This is not ion-induced after-pulsing, as that occurs on a timescale of a few $\mu$s \cite{Faham:2014gyi}. Rather, it seems to be a more delayed photocathode after-pulse effect. We are not aware of other reports or studies of this effect, though it appears to also be evident in Fig 12 of Ref. \cite{LUX:2020vbj}. For the present work, suffice it to say that for excitation pulses greater than 500 photons detected, the $50-100~\mu$s data point is excluded from the fit of $d_p(t)$.

In this experimental configuration, we systematically removed each material suspected of fluorescing: PTFE was replaced with Kapton and then with aluminum, and the R8778 sensor was replaced with S13371 sensors, and vice-versa, and the stainless steel electrode was removed on suspicion of surface oxide fluorescence. We also varied the height of the liquid/vapor interface, including removing it altogether by overfilling the apparatus, and repeated the studies of varying the applied electric field. None of these changes had any effect on the resulting power law $d_p(t)$, other than to decrease the total photon detection efficiency. Following these tests, the dominant source of $d_p(t)$ could be narrowed to two remaining suspects: (1) the xenon (or impurities therein), which seems reasonable considering measurements of other condensed noble gases \cite{PhysRevA.67.062716}; or (2) the photosensors. The primary common element between the S13371 and R8778 photosensors is the quartz window.

\subsection{Measurement of quartz fluorescence}
In order to test the hypothesis that fluorescence of the quartz window causes delayed photons, we made the same series of measurements with vacuum in the active region instead of liquid xenon. Gaseous xenon was initially used to thermalize the entire testbed for a period of several days, after which the xenon was removed and the active region was evacuated to a pressure of $1\times10^{-6}$~mBar. This low pressure was maintained by constant vacuum pumping during the measurement. The temperature of the testbed was then maintained at $T=(-99 \pm 1)$~$^{\circ}$C.

Cherenkov photons from several MeV background beta decays in or on the surface of the PMT window were used as a source of UV photon pulses. The Cherenkov signal is directional. {\color{black} An average excitation pulse $\bar{a} = 329$ photons detected by the PMT were the largest such signals we could obtain. The largest $\bar{a}$ registered by the S13371 SiPM, regardless of trigger configuration, was about $\times0.1$ that size, so they were not used for this measurement.} The Cherenkov signal decays faster than xenon scintillation, and has a continuous spectrum that rises toward shorter wavelengths as $dN/dx \sim 1/\lambda$ (cf. Eq. 34.44 of \cite{ParticleDataGroup:2024cfk}). The PMT response cuts off at $\lambda=160$~nm. Considering the response of the PMT extends to about 650~nm, we calculate that nearly half of the detected photons are in the range $160 < \lambda < 250$~nm. The results are shown in Fig. \ref{fig:vac}. This measurement confirms that quartz is a source of UV-induced fluorescence with a magnitude and time profile similar to the observed $d_p(t)$ in liquid xenon TPCs. 


\subsection{Quartz is the dominant source of fluorescence}
\begin{figure}[ht]
    \centering
    \includegraphics[width=0.49\textwidth]{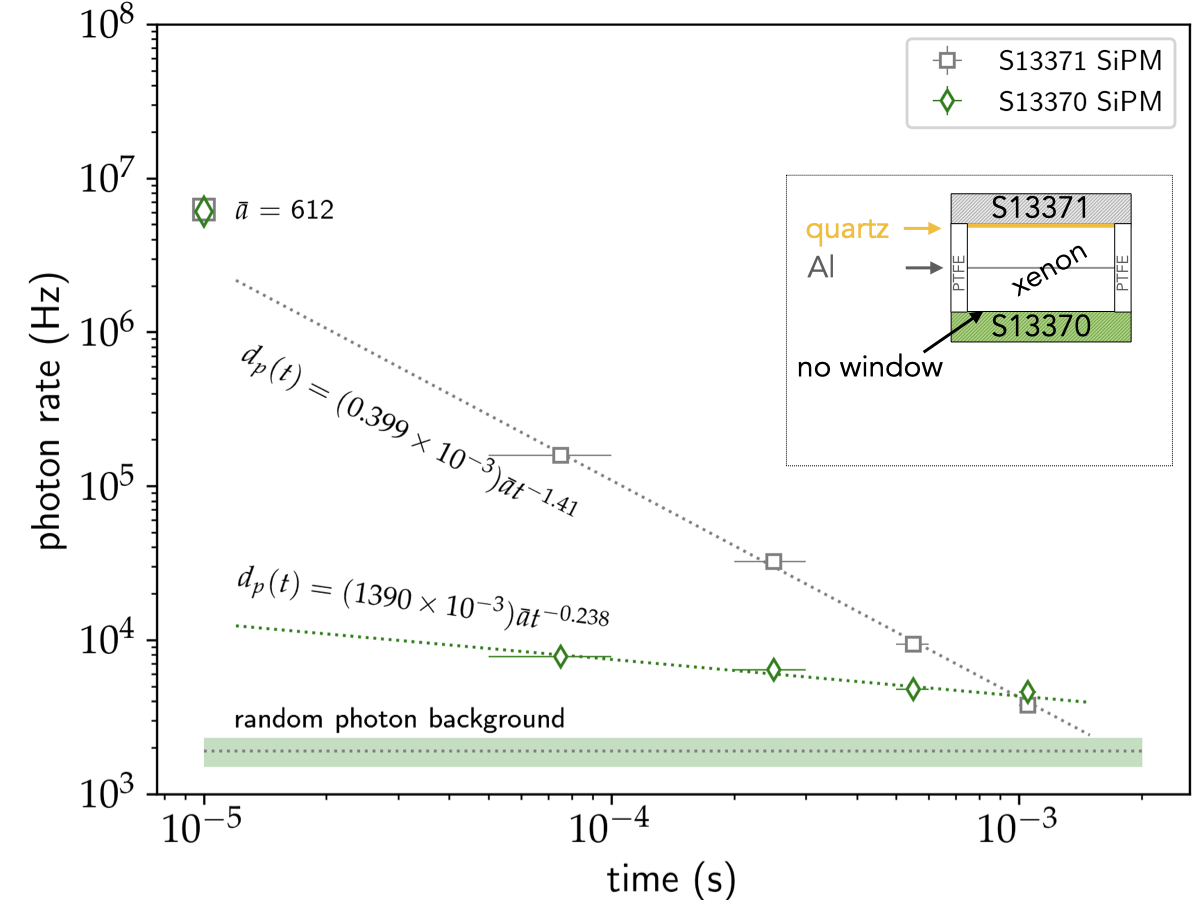}
    \caption{Measurements of the delayed photon rate following an excitation pulse $\bar{a}$ of UV photons from Xenon scintillation. The experimental configuration is indicated inset, featuring two symmetric, optically isolated regions, one with a quartz window and one without.}
   \label{fig:no-window}
\end{figure}

In order to identify quartz fluorescence as the primary source of delayed photons induced by xenon scintillation, we adapted the experimental configuration as shown in Fig. \ref{fig:no-window}: the active region was divided with a piece of aluminum, in order to optically isolate the upper and lower halves. The PMT was replaced with an array of four, windowless Hamamatsu S13370 SiPM. Of the four sensors, two exhibited an order of magnitude greater leakage current and dark count rate, and so were left unbiased. Of the remaining two, one exhibited $53\%$ higher dark count rate. We show data from the lower rate sensor, but note that nearly identical results are obtained with the higher rate sensor. A single channel of the twelve S13371 sensors was used as a comparison in this case. The active region was completely filled with liquid xenon, and a flow-through \textsuperscript{220}Rn alpha particle source was used to provide xenon scintillation pulses. Because of the optical isolation, triggering on a pulse from the S13371 sensor was accompanied by random (mostly zero) photon counts in the S13370 sensor, and vice versa. 

The two data sets were acquired separately and are shown together in Fig. \ref{fig:no-window}. The measured random photon background rate in the single S13371 sensor is an order of magnitude smaller than that of the S13370, and is below the axis range. The S13370 sensor without a quartz window measures $d_p(t)$ reduced by a factor $\times20$ in the $50-100~\mu$s bin. A new, residual component of delayed photon noise observed by this sensor appears to follow a power law with a much shallower slope $k\simeq 1/4$.  We know from previous investigation \cite{Sorensen:2024idm} that the power law measured by the S13371 is unbroken to at least 5 ms, which would be at odds with interpreting this new component as an effect of the xenon or its impurities. We therefore suspect this residual component is due to fluorescence of the ceramic package of the S13370 sensor. Ceramic is known to fluoresce \cite{Koo:10}. Visually, it is clear that the S13370 uses a different ceramic than the S13371. Considering the drawings on the datasheets, it also has a factor $\times4$ more ceramic facing the xenon.


\section{Discussion}
We have identified the dominant component of $t<5$~ms timescale delayed photon noise in liquid xenon particle detectors as UV-stimulated fluorescence from quartz windows in the photosensors. Although photoluminescence of glass due to impurities is well-known \cite{Jedamzik:2017}, we are not aware of any other measurements of the time dependence of quartz glass fluorescence on timescales $\lesssim1$~second. Power law fluorescence is generally attributed to charge-trapping on point defects \cite{Huntley_2006}. The fluorescence is often referred to as phosphorescence when the lifetime extends into the ms regime \cite{Jedamzik:2017} (as in the present case). A very high purity quartz called VIOSIL manufactured by Shin-Etsu \cite{Shinetsu} appears to have no measured fluorescence, in contrast to fused quartz.  Early work \cite{Coop:62} found a power law decay on a scale of minutes following exposure to intense source of ultraviolet radiation. More recent studies of electron-induced fluorescence in PMT windows \cite{Viehmann:1975,Osterman:2005} were motivated by excessive dark counts observed in PMTs on space flight experiments, and found multiple exponential components to the fluorescence on timescales $\gtrsim1$~minute.

A key question remains: is the ms-scale quartz fluorescence still dominant at later times $5\lesssim t \lesssim 1000$~ms, the regime of relevance for dark matter search experiments? This regime is difficult to probe with a high-rate test bed at the surface. Conversely, the ms regime is difficult to probe in the underground dark matter search experiments, due to the time structure of their signals extending to $t\simeq1$~ms.

\begin{table}[h]
{ \footnotesize
\begin{center}
\renewcommand{\arraystretch}{1.2}
\caption{Summary of power law parameter measurements}
\label{tab:alpha}
\begin{ruledtabular}
\begin{tabular}{llllll}

\textbf{Exp.}	       &	
\textbf{Sensor}	       &	
\textbf{Excitation}    &
\textbf{$\alpha \times10^{3}$}	&	
\textbf{$k$}	 &	
\textbf{Range (s)}	\\ 
\colrule
LBL & S13371 & Xe& 2.30 & -1.08 & $10^{-5} - 10^{-3}$	\\ 
LBL & R8778 & Xe& .200 & -1.31 & $10^{-5} - 10^{-3}$	\\ 
LBL & R8778 & Cherenkov & 2.22 &-1.03 & $10^{-5} - 10^{-3}$	\\ 
LBL & S13371 & Xe & .399 &	-1.41 & $10^{-5} - 10^{-3}$	\\ 
LBL$^a$ & S13370 & Xe & 1390 & -.238 & $10^{-5} - 10^{-3}$	\\ 
LUX$^{b}$ & R8778 & Xe & 1.00 & -.500 & $10^{-3} - 1$	\\ 
LZ$^{c}$ & R11410 & Xe, $\bar{a}=10^{6.5}$ & 2.64 & -.953 & $10^{-2} - 1$	\\ 
LZ$^{c}$ & R11410 & Xe, $\bar{a}=10^{5.5}$ & 22.1 & -.542 & $10^{-2} - 1$	\\ 
LZ$^{c}$ & R11410 & Xe, $\bar{a}=$~all & 0.42 & -.1.3 & $10^{-2} - 1$	\\ 

\end{tabular}
\end{ruledtabular}
\begin{flushleft} $^{a}$No quartz; $^{b}$Ref. \cite{LUX:2020vbj}, our fit; $^{c}$Ref. \cite{Anderson-thesis}. \end{flushleft}
\end{center}
} 
\end{table}

Limited data are available for direct comparison of the power law parameters $\alpha$ and $k$. Within the same experiment, $\alpha$ can vary by an order of magnitude, and $k$ by nearly a factor of $\times2$. This systematic uncertainty is due to the combination of (1) event rate and (2) event selection: 
\begin{enumerate}
\item higher event rates lead to shorter inter-event times, thus a larger measured $\alpha$ for any particular event, due to the receding power law tail. This should tend to flatten the time profile of $d_p(t)$, resulting in a larger $k$. The effect is evident in our measurements utilizing the S13371 sensors, with data shown in Fig. \ref{fig:xe} having a factor $\times2$ higher source rate than data shown in Fig. \ref{fig:no-window}. 
\item similarly, if smaller events are selected to comprise the excitation pulse $\bar{a}$, then larger events in the data stream present larger power law tails upon which $d_p(t)$ is measured. This effect moves in the same direction as (1), and is also evident in the two LZ fits cited. A solution to this systematic uncertainty is described in Sec. 5.7.1 of \cite{Anderson-thesis}, in which a continuous data stream of events are simultaneously fit for their respective $d_p(t)$ (last row in Table \ref{tab:alpha}).
\end{enumerate}

We emphasize that while a more precise characterization of $\alpha$ and $k$ are worthy of further study, the key point for future dark matter searches, and therefore of the present work, is simply to identify the cause of $d_p(t)$ and eliminate it. To this end, we offer suggestive if inconclusive evidence that the ms-scale quartz fluorescence is consistent with the $>10$~ms-scale photon noise previously measured in LUX \cite{LUX:2020vbj} and LZ \cite{Anderson-thesis}. 

Several follow-on studies are needed: 
\begin{enumerate}
\item the dark matter search experiments could attempt to correlate measured $\bar{a}$ with $d_p(t)$ on a per-sensor basis. If quartz fluorescence is the dominant source, this correlation should exist. Anecdotally, this analysis has been attempted, and no effect has been seen. Such a null result should be expected in the absence of careful event selection. Specifically, events with large $\bar{a}$ (relative to a stream of preceding events, and long delay times to those preceding events, might be able to show a correlation. Otherwise, we expect the correlation could be washed out by the power law 
nature of the effect. {\color{black} It is possible that total internal reflection at the PMT quartz-vacuum interface limits the magnitude of the correlation.}
\item additional small scale test bed studies are needed, with a lower background counting rate and a cascade trigger out to longer times. Bare silicon SiPM sensors should be studied, to mitigate the effect of ceramic fluorescence. If such an experimental configuration can be realized, it would be able to probe possible fluorescence from impurities within the xenon or perhaps from the xenon itself. 

\end{enumerate}

This investigation is critical to the design choices of future experiments, such as the proposed XLZD experiment \cite{Aalbers:2022dzr}. For example, future R\&D should focus on characterizing the fluorescence from the R11410 PMT favored by the leading dark matter search experiments, and on minimizing the UV-induced fluorescence from window materials such as quartz, MgF, and others. Assuming quartz fluorescence dominates $d_p(t)$ at later times $5\lesssim t \lesssim 1000$~ms, improvements in window materials could lead to a much reduced delayed noise rate in future experiments. Such a result would favorably improve the ratio of signal to AC background, leading to an increase in discovery potential. Alternatively, silicon photomultipliers do not strictly need a window at all, other than to protect the sensor face. However, silicon-based sensors have other challenges, such as external crosstalk~\cite{Gibbons:2023iux} and higher dark count rates even for optimized device designs \cite{Sakamoto:2023ond}.

\begin{acknowledgments}
We thank Shingo Kazama, Tom Shutt and Jingke Xu for interesting discussions and comments. This work was supported by the U.S. Department of Energy (DOE) Office of Science, Office of
High Energy Physics under contract DE-AC02-05CH11231. PS additionally acknowledges support from the DOE Early Career Research Program.

\end{acknowledgments}

\bibliography{phtrain}

\end{document}